\documentclass[sigplan,screen]{acmart}
\usepackage{appendix}
\AtBeginDocument{%
  \providecommand\BibTeX{{%
    \normalfont B\kern-0.5em{\scshape i\kern-0.25em b}\kern-0.8em\TeX}}}

\setcopyright{acmcopyright}
\copyrightyear{2021}
\acmYear{2021}
\acmDOI{10.1145/1122445.1122456}

\acmConference[Conference '21]{Conference '21}{Mon 11 - Fri 15 October, 2021}{Bari, Italy}
\acmBooktitle{Conference '21}
\acmPrice{15.00}
\acmISBN{978-1-4503-XXXX-X/18/06}

\begin{document}

\title{A Survey on the Techniques and Tools for Automated Requirements Elicitation and Analysis of Mobile Apps}


\author{Chong Wang}
\affiliation{
  \institution{Wuhan University}
  \city{WuHan}
  \country{China}}

\author{Haoning Wu}
\affiliation{
  \institution{Wuhan University}
 \city{WuHan}
  \country{China}
}

\author{Peng Liang}
\affiliation{
  \institution{Wuhan University}
  \city{WuHan}
  \country{China}
}

\author{Maya Daneva}
\affiliation{
  \institution{University of Twente}
  \city{Enschede}
  \country{the Netherlands}
}

\author{Marten van Sinderen}
\affiliation{
  \institution{University of Twente}
  \city{Enschede}
  \country{the Netherlands}
}


\begin{abstract}
\textbf{[Background:]} Research on automated requirements elicitation and analysis of mobile apps employed lots of techniques and tools proposed by RE researchers and practitioners. However, little is known about the characteristics of these techniques and tools as well as the RE tasks in requirements elicitation and analysis that got supported with the help of respective techniques and tools.
\textbf{[Aims:]}The goal of this paper is to investigate the state-of-the-art of the techniques and tools used in automated requirements elicitation and analysis of mobile apps.
%and recommend the best techniques and tools for researchers to handle different automated requirements elicitation and analysis of mobile apps tasks .
\textbf{[Method:]}We carried out a systematic mapping study by following the guidelines of Kitchenham et al. 
\textbf{[Results:]} Based on 73 selected papers, we found the most frequently used techniques -  semi-automatic techniques, and the main characteristics of the tools - open-sourced and non-self-developed tools for requirements analysis and text pre-processing. Plus, the most three investigated RE tasks are requirements analysis, mining and classification.\textbf{[Conclusions:]} Our most important conclusions are: (1) there is a growth in the use of techniques and tools in automated requirements elicitation and analysis of mobile apps, (2) semi-automatic techniques are mainly used in the publications on this research topic, (3) requirements analysis, mining and classification are the top three RE tasks with the support of automatic techniques and tools, and (4) the most popular tools are open-sourced and non-self-developed, and they are mainly used in requirements analysis and text processing. 

\end{abstract}




\maketitle

\section{Introduction}
With the increasing popularity of mobile devices, the mobile application market is expanding to involve more mobile applications (app for short). This inevitably results in the massive app data to be processed and analyzed for different RE purposes of mobile apps. Since it is a time-consuming or even impossible task to handle such a large number of app data by manual, various types of techniques and tools have been successfully conducted on the app datasets in the literature to facilitate the automated requirements elicitation and analysis of mobile apps. Considering the five essential RE activities, requirements elicitation and analysis have been reported as the most two investigated RE activities based on app datasets~\cite{9}. As we know, some studies explored tools and techniques used in RE activities of mobile apps. However, there is no systematic mapping study(SMS) on this research topic. Therefore, this paper aims to investigate the state-of-the-art of the techniques and tools used in automated requirements elicitation and analysis of mobile apps, by following the guidelines of Kitchenham et al.\cite{8}

The main contributions of this SMS are as follows. First, it contributes to the emerging literature on the techniques and tools used in automated requirements elicitation and analysis of mobile apps. Specifically, it consolidates the empirical research published on the sources of automatic/semi-automatic techniques and tools that are employed in and beneficial for the aforementioned two RE activities. Second, it indicates the main characteristics of both techniques and tools that used in the automated requirements elicitation and analysis of mobile apps.

This rest of this paper is organized as follows: Section 2 discusses the related work. Section 3 explains the process of conducting a systematic mapping study. Section 4 and Section 5 present the research questions and the research process specified in this SMS. Section 6 and 7 are the preliminary results of this SMS and our discussion on these results, respectively. Section 8 discusses the limitations, followed by the conclusions and future work in Section 9. 

\section{Related Work}
Several reviews and systematic mapping studies investigated techniques and/or tools used for apps. In\cite{1}, the researchers analyzed the information, training process, and evaluation metrics related to the selected application algorithms. They found that Naive Bayes, Decision Trees, and Natural Language Processing algorithms were the three most commonly used. \cite{2} examines the status of automated requirements elicitation tools. They mainly concentrated on scopes and degree of automation of these tools. \cite{3} reported 24 types of ML-based approaches in a systematic review, in order to identify and classify NFRs. Whereas, \cite{4} is a systematic mapping study on sentiment analysis by using SVM (support vector machine). They serve the scholars and researchers to analyze the latest work of sentiment analysis with SVM as well as provide them a baseline for future trends and comparisons. In \cite{5}, the authors explored and compared the accuracy of classifying functional and non-functional requirements with random forest algorithm and gradient enhancement algorithm. \cite{6} investigated how to accurately and automatically classify requirements into functional and non-functional requirements with supervised machine learning algorithms. \cite{7} identified existing methods and tools for traceability between software architecture and source code, as well as empirical evidence of these methods, their benefits in relation to software architecture understanding, and issues and challenges.

To best of our knowledge, seldom systematic mapping studies or surveys investigate an overview of techniques and tools for automated requirements elicitation and analysis of mobile Apps.

\section{Systematic Mapping Study Process}

An SMS is a well-defined and rigorous method for identifying, evaluating, and interpreting all available research relevant to a particular research question, topic area, or phenomenon of interest, focuses on providing a wide overview of a domain, identifying research evidence on a topic, and presenting mainly quantitative results.

This study was conducted by following the guidelines of Kitchenham et al.\cite{8}, which is composed of three main stages:

1. Review preparing ,which aims at determining the main content of the review and planning a review protocol.

2. Review implementing ,which aims at accomplishing the review by executing the prepared review protocol from the previous phase.

3. Review discussing ,which aims at presenting the results and discussing the results according to our research directions needed.

The purpose of this review is based on the goal-question-measure approach developed by the SMS mentioned above.Analyze primary studies for the purpose of exploration and analysis with respect to the tools and techniques used in the mobile application research from the point of view of researchers and practitioners in the context of requirements elicitation and analysis.

In order to obtain a detailed and comprehensive result, the topic of our SLR can be divided into 3 RQs. Specifically, we screen all articles involving mobile application requirements analysis and requirements elicitation between 2012 and 2020, and filter those that meet our requirements, and we find that by analyzing articles within this time period, we can meet our research needs well. 

\section{Review preparing}

The first phase of the SMS is to perform the pre-review activities for the system review. The review plan involves the steps of goal and research questions (see Section ~\ref{4.1}), inclusion and exclusion criteria (see Section ~\ref{4.2}), search strategy (see Section ~\ref{4.3}).

\begin{table*} [t]
  \caption{Inclusion and exclusion criteria in our SMS }
  \label{criteria}
  \begin{tabular}{p{8.5cm}p{8.5cm}}
    \toprule
    Inclusion Criteria (IC) & Exclusion Criteria (EC)\\
    \midrule
     IC1: The paper directly relates to the topic of requirements elicitation and analysis of mobile applications. & EC1: The paper addresses the design and/or implementation of a specific mobile application.\\
     IC2: The paper addresses all the research questions. & EC2: The paper does not aim to requirements elicitation and analysis of mobile apps.\\
     IC3: The paper aims to functional requirements and quality requirements specified in ISO 25010. & EC3: The dataset is not directly used for the RE purpose of mobile apps.\\
     IC4: The paper is published in a peer-reviewed journal, conference, or workshop. & EC4: The paper is only a research plan or literature review. \\
     IC5: Tools or techniques are employed for the automated RE activities. & EC5: The full paper is not written in English.\\
     IC6: The paper uses automatic or semi-automatic methods for the automated RE activities of mobile apps. & EC6: The full paper is not available for download.\\
     ~ & EC7: The full paper only uses manual studying methods. \\
    \bottomrule
  \end{tabular}
\end{table*}

\subsection{Goal and Research Questions}\label{4.1}
There are several primary studies on techniques and tools used in automated requirements activities for mobile applications (such as [1,3,7]),
but so far no systematic mapping study has been undertaken. 

This SMS is conducted as the first systematic mapping study to explore the state-of-the-art of algorithms and tools supporting automated requirements elicitation and analysis for mobile applications in a thorough and unbiased manner.
% Moreover, this SMS is a first step towards dra)wing more general conclusions. Accordingly, 
To researchers, this SMS provides a resource to determine the gaps and pointers for further research in this area. To practitioners, the SMS provides a pool of effective and popular algorithms to support automated RE activities for mobile apps.

\textbf{\textit{RQ1: What techniques have been used in automated requirements elicitation and analysis of mobile apps?}} This RQ investigates the types of techniques used for automated requirements elicitation and analysis of mobile apps. We answer this RQ by examining the types of automatic techniques, the automation degree of techniques, as well as whether these techniques were directly reused or proposed by the authors.  

%the automaticity of the specific algorithms used in the different aspects of the study identified for the different apps, the categories to which they belong, and whether some degree of improvement and innovation has been made for the corresponding problems.

\textbf{\textit{RQ2: In what tasks of requirements elicitation and analysis of mobile apps have the techniques been applied to?}} In this RQ, we examine the purpose of employing automatic techniques in requirements elicitation and analysis or mobile apps. This RQ provides concrete RE tasks in requirements elicitation and analysis activities that got supported from automatic techniques. 
%We classify these purposes into two categories according to theories related to requirements engineering, requirements analysis, and requirements acquisition. Answering this question helps us to understand the usefulness of the algorithms used in requirements engineering-oriented mobile application research for RE activities

\textbf{\textit{RQ3: What tools have been used to facilitate automated requirements elicitation and analysis of mobile apps?}} This RQ explores the characteristics of the tools to support automated requirements acquisition and analysis of mobile apps. We answer this RQ by examining the types and open accessibility of these tools, in order to suggest popular and effective tools for the RE purposes of mobile apps. 

%\subsection{specifying query questions}
%The research problem identification step is considered to be one of the most important parts of the SMS, as the re-search problem drives the entire system mapping study.
%The search problem is the driving force of the entire system mapping study. The research questions chosen to achieve the objectives of our study are listed below:

\subsection{Inclusion/Exclusion Criteria}\label{4.2}
Inclusion and exclusion criteria are crucial to retrieve relevant literature and construct a initial pool of publications for manual filtering. Table 1 lists the inclusion and exclusion criteria in this SMS. 

%. were used to determine inclusion and exclusion criteria are used to determine the suitability of a publication (article) and to make a decision whether to accept or reject a particular publication for inclusion in SMS.The inclusion and exclusion criteria for this report are listed below the inclusion and exclusion criteria for SMS are listed below:

\subsection{Search Strategy}\label{4.3}
\label{search_words}
%The search strategy step of SMS is intended to provide the basis for a comprehensive and unbiased collection of research in the literature.
It is important to specify the data sources of the research and to develop a search strategy so that each relevant bibliographic citation has a good chance of appearing in the search results. 
%This is because it affects the quality and completeness of the retrieved research.
We searched two digital libraries, i.e., IEEE Xplore Digital Library and Scopus, to locate literature for this SMS. 
%selected the IEEE database and the Scopus database, which intersect in the specific data acquisition, by further filtering. These two databases can cover the vast majority of papers needed for our study.

The search string comprises keywords derived from the research topic and Boolean logic for performing efficient searching. 
%The following search string is created to search for relevant publications:
First, since this SMS aims to the RE of mobile apps, the term `mobile apps' and its alias (such as apps or mobile application) have to be included in the search string. Second, the term `requirements' and its broader concepts (i.e., `feature' and RE) are suggested as the keywords, in order to (a) concentrate on two RE activities of mobile apps, i.e., requirements elicitation and analysis, and (b) to cover publications that did not explicitly mention `elicitation' or `analysis' in the full text but reporting these two RE activities. Finally, we scoped the time period of the related publication from January 2008 to December 2020 because (a) both the Apple App Store and Google Play were launched in 2008; and (b) our automatic search was conducted in January 2021, and the search output published in 2021 is incomplete for this SMS. 
%be expressed in various ways in articles, such as app or mobile application, but some articles will use specific programs or a particular research area, and we exclude them unless they are related to the overall study of requirements engineering.
% In order to further analyze the specific requirements engineering involved in articles oriented towards requirements engineering analysis, we used ("Abstract": requirements) OR "Abstract": feature) OR "Abstract": RE), and any articles that mentioned requirements activities, we retrieved them first.
% Through experimentation and selection and further optimization, we determined the retrieval string to be
The finalized search string is presented below.

\textit{TITLE:( "app" OR "mobile application" ) AND Abstract:( "requirements" OR "feature" OR "re" ) ) AND The publication time range is limited to January 2008 to December 2020}

\section{Review Implementing}
%The second phase of the SMS approach is to conduct an SMS. t
The steps of conducting an SMS include search and selection of primary studie in Section ~\ref{5.1}, data extraction in Section ~\ref{5.3}.

\subsection{Search and Selection of Primary Studies}\label{5.1}
By implementing the search string defined in Section~\ref{search_words} in IEEE Xplore and Scopus respectively, 762 papers were returned from IEEE Xplore and 2065 papers from Scopus that form the raw set of papers for our SMS. Before filtering the publications manually, 299 duplicates out of the 2827 papers were removed by using the Pandas library in Python. As a result, 2528 papers were retained as the starting point for the three-round selection process.

%To build the initial article library, we searched the Scopus database and the IEEE database using the search terms mentioned in the previous section, respectively, and the above databases were used as the database search sources for this SMS. Finally, after searching, IEEE got 762 articles, Scopus got 2065 articles.

%After obtaining this initial body of literature, we reviewed each of the papers.

% In our comparison experiments we found that there is a degree of overlap in the scope of the scopus database and the IEEE database. And our specific experiments require non-duplicated data for filtering. So we filtered and cleaned the sample data by the tool made by the pandas library in python. Finally, we got the complete non-duplicated data set required for the experiment. The final result is 2528 de-duplicated papers

As shown in Figure 1, the authors first reviewed the titles of 2528 papers. After excluding 2100 papers, 428 papers were sent to the second round selection by abstract. This results in excluding 301 papers. By reviewing the full text of the remaining 127 papers in the third-round selection process, we further excluded 54 papers and got a final set of 73 papers as the primary studies for this SMS. The list of the 73 primary studies is available at \emph{Google Doc}\footnote{https://docs.google.com/document/d/1M75KDvMlZtKV1py4u9zuHRPLIh0XV675OAj\\93BTJ9N4/edit?usp=sharing} 

%Firstly, we selected the collection of papers that might be relevant based on whether the title was relevant to our research topic. 

%Secondly,From the set of papers filtered by title, we again filtered by abstract. to find if the abstract contains all the keywords we need.
\begin{figure}[h]\label{figureone}
\label{process}
  \centering
  \includegraphics[width=\linewidth]{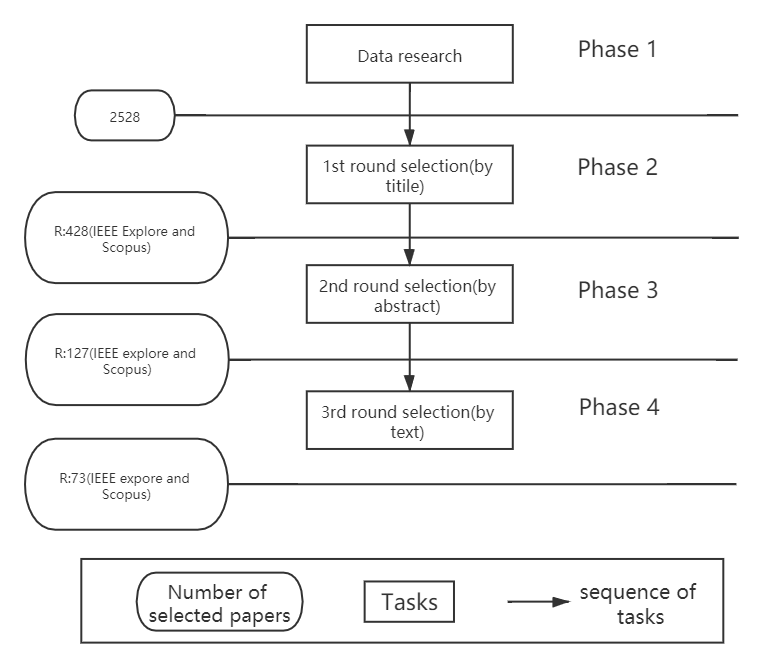}
  \caption{Three-round selection of primary studies.}

\end{figure}

The three-round selection process was conducted by the first three authors. The first author performed the 1st and 2nd rounds selection, by leaving the uncertain papers to the next round. During the 3rd round selection, when the first author was not confident in the exclusion or inclusion of certain papers, he discussed with the second author to get a consensus. As soon as the 3rd round selection was completed by the first author, the second and the third authors randomly selected 20\% of these included papers and gave them careful reading for cross-validating the included papers, and got 100\% agreement on these selected papers.

%the first and the second authors participated in the first and the second round the Thirdly,After the first two rounds of screening, we performed further screening by reading through all the papers obtained. After obtaining the collection of papers, we reviewed each paper. The second author independently checked all citations and in case of doubt, we discussed the citations in detail for detailed discussion. Finally, both authors read all selected publications in detail.

%During the database search, 2528 papers were retrieved from the two databases. 428 papers were included after the 1st round selection and 127 papers were retained after the 2nd round selection and 73 papers were retained after the 3rd round selection. Furthermore, every papers were identified reviewed by second author, this led to a final set of 73 papers in this SMS.

\subsection{Data Extraction}\label{5.3}
In this SMS, we used the data extraction form in Table 2 to extract data from the 73 primary studies for answering the three RQs in Section ~\ref{4.1}. In particular, we extracted publication title and year of each selected study to give a overview of the 73 primary studies.

To answer the three RQs defined in Section 3.1, we used the data extraction form in Table 2 to extract data items from the 44 primary studies. Regarding the demographics of the primary studies, the data items are directly detected in the publication information of the selected studies.

\begin{table} [h]\label{tbtwo}
  \caption{Data Extraction Form}
  \label{extraction}
  \begin{tabular}{p{1.8cm}p{5.5cm}}
    \toprule
    Relevant RQ & Data Items\\
    \midrule
     RQ1 & Techniques or algorithms used for automated requirements elicitation and analysis\\
       ~ & The automation of the techniques or algorithms \\
       ~ & The types of the techniques or algorithms \\
    RQ2 & Requirements activity involved in the selected studies\\
    RQ3  & Name and usage of the tool for requirements elicitation and analysis\\
      ~ & The main function of the tools \\
      ~ & Whether the tool is self-developed, and whether the tool is open-sourced. \\
    \bottomrule
  \end{tabular}
\end{table}

\section{Results}
This section collects and synthesised the data extracted from the 72 primary studies, gives an overview of these primary studies, and provides answers to the aforementioned three RQs.  

%This section presents the SMS results of our approach to techniques and tools for automated requirements eliciation and analysis of mobile apps. It aims to collect and summarize the results of the included major studies.It aims to collect and summarize the results of the included major studies and to provide a comprehensive and orthogonal classification of the results of these studies. We organized the presentation of the SMS results according to the order of the three RQs mentioned in Section 4.1.The order of the five sections is as follows: Search and Selection Results (Section 6.1),Overall data situation (Section 6.2), RQ1 (Section 6.3), RQ2 (Section 6.4), RQ3 ( Section 6.5).

\subsection{Demographics of Primary Studies}
Figure~\ref{type} shows the publication types of the included 73 studies. We observed that around 68.5\% (50 out of the 73 studies) are conference papers. Workshop and journal publications account for 12 and 11 studies, respectively.  
%the 73 s  illustrates the three types of selected studies: conference, journal, workshop. Most of the studies were published in conferences(50 out of 73, 68.5$\%$), workshop and journal are closer, workshop(12 out of 73, 16.4$\%$) and journal(11 out of 73, 15.1$\%$).

\begin{figure}[h]\label{type}
  \centering
  \includegraphics[width=\linewidth]{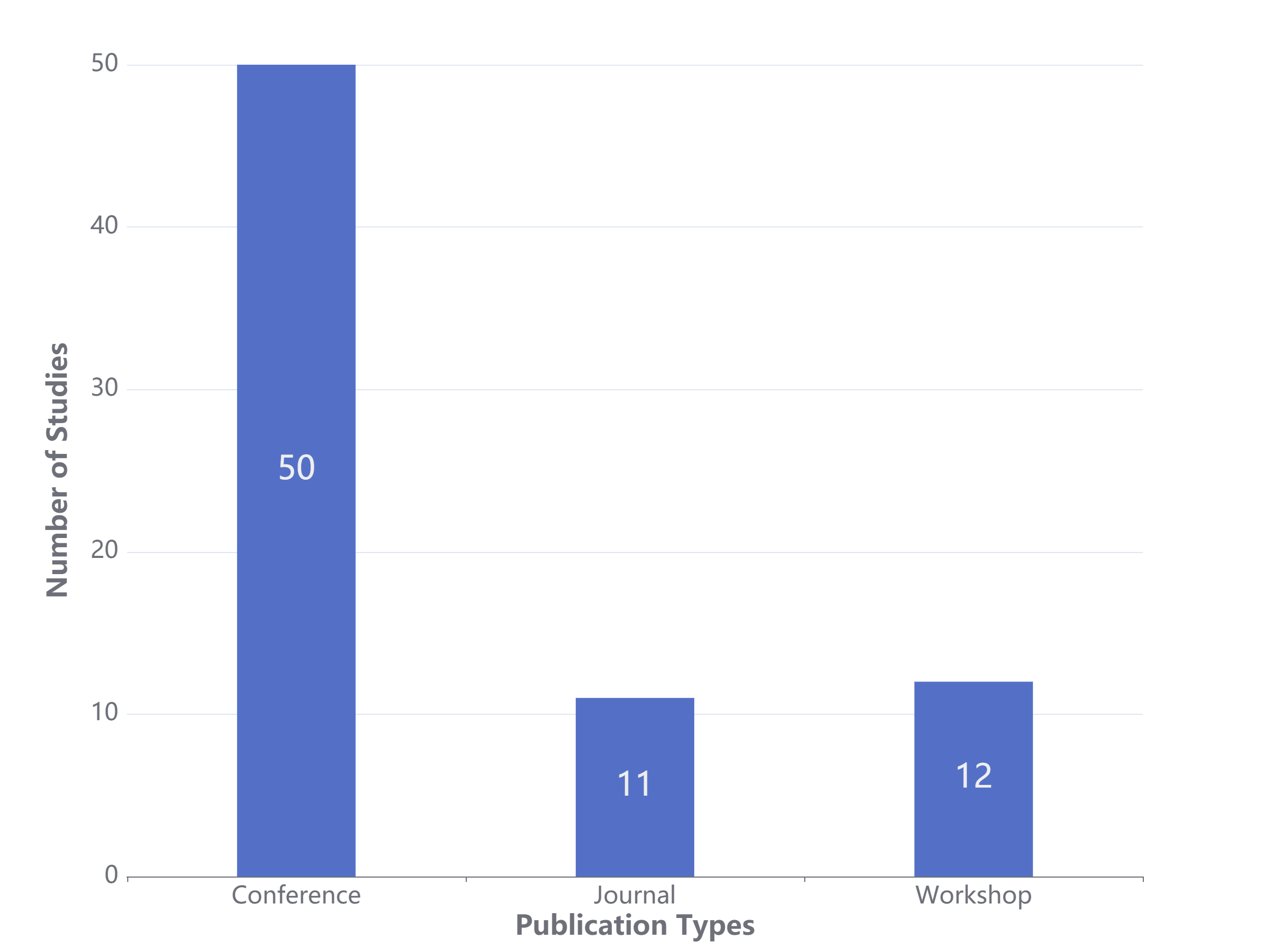}
  \caption{Number of studies over the three publication types.}
  \label{type}
\end{figure}

The distribution of the selected studies over years is shown in Figure~\ref{year}. We found that the research employing techniques and tools for automated requirements elicitation and analysis of mobile apps becomes more and more popular since 2017. Especially, the number of the selected studies in 2017 is four times of the ones published in 2016, and the number of selected publications reached a peak in 2019, accounting for 20 studies. In addition, it is surprising to find that the number of literature published in 2020 was jumping to 12 studies.   

\begin{figure}[h]\label{year}
  \centering
  \includegraphics[width=\linewidth]{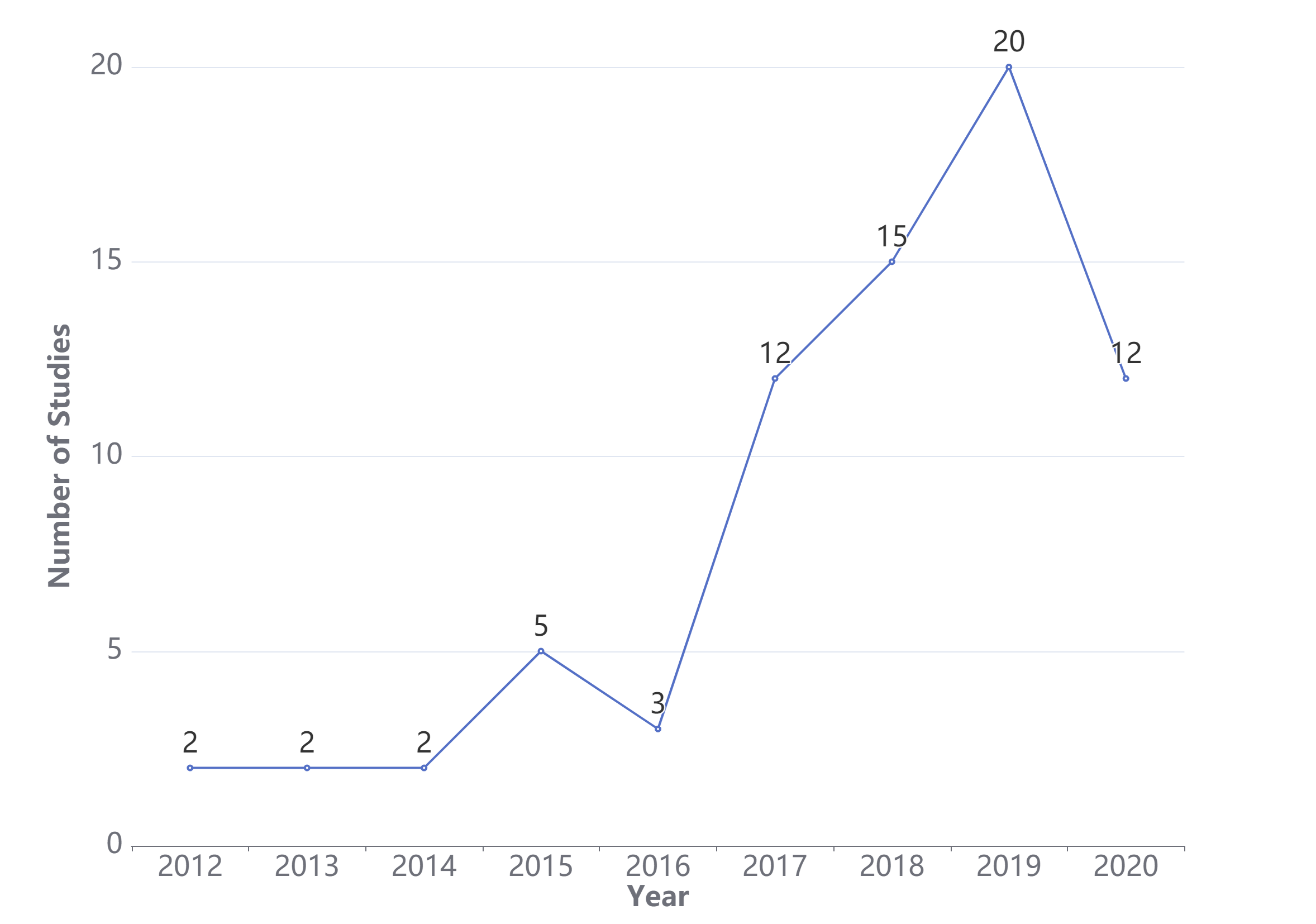}
  \caption{Number of studies over time period (2012–2020).}
  \label{year}
\end{figure}

\subsection{Answer to RQ1}\label{6.2}
The automated requirements elicitation and analysis of mobile apps needs support from automatic techniques, methods, and algorithms. The techniques used in different studies may vary, even for the same RE purpose. In this section, we investigated the techniques employed in the automated requirements elicitation and analysis from the following three aspects, i.e., the automation degree of the techniques, the types of automatic techniques, and the novelty of the techniques. 

\subsubsection{The Automation Degree of Techniques}
%The degree of automation of techniques affects the overall execution process and execution efficiency of the study. 
In our SMS, the automation degree of techniques is evaluated from the perspectives of automated or semi-automated. Automated means that the in a certain study, whole dataset of app data is processed automatically with the technique. Whereas, semi-automated refers to situation that both manual and automatic techniques are applied in a certain study to process the app data. 

\begin{figure}[h]
  \centering
  \includegraphics[width=\linewidth]{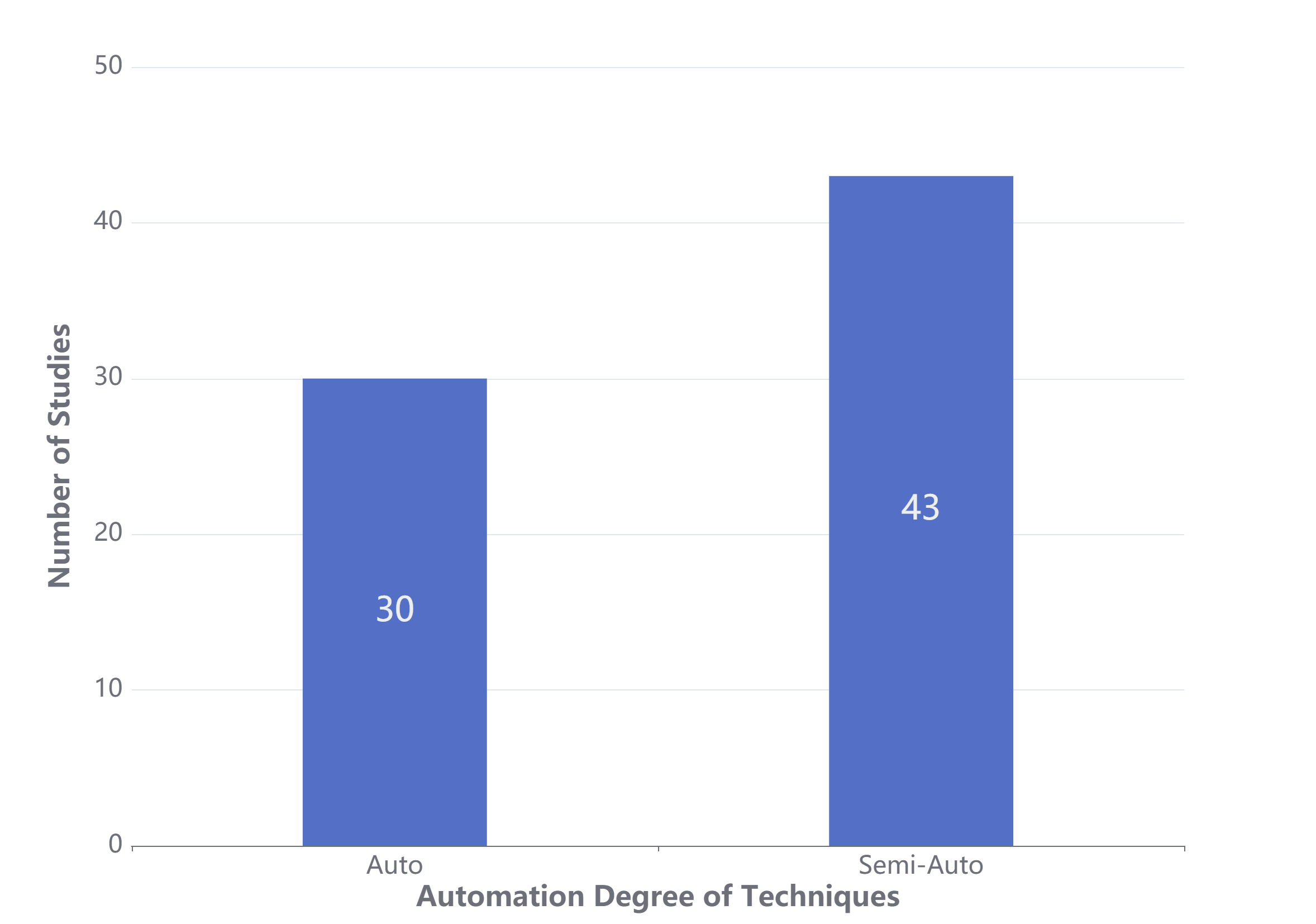}
  \caption{Distribution of the selected studies over the automation degree of techniques.}
  \label{auto}
\end{figure}

Figure~\ref{auto} shows the distribution of the 73 studies over the automation degree of techniques. We found that 30 out of 73 selected studies employed pure automatic techniques for automated requirements elicitation and analysis. For example, S31 proposed the method named CHANGEADVISOR to extract and classify the informative sentences of app reviews, in which NLP (Natural Language Processing) and SA (Sentiment Analysis) techniques were used in processing the app datasets. In S43, the researchers directly employed the random forest algorithm to automatically extract features from user reviews to model developers' response behavior. Whereas, in around 59\% of the selected studies (43 studies), the automatic techniques were applied with the supplement of manual methods. In S14, for example, the app reviews were first manually labelled as either informative or non-informative to construct the truth set, before using the machine learning algorithms for the automatic classification of app reviews. 
%In [S47], researchers sent emails to different developers with two research questions about their research directions as part of the initial database as part of the initial database.
In [S63], sample reviews were manually classified into two categories, i.e., functional and non-functional requirements, in order to automatically extract keywords from the test set of app reviews.

\subsubsection{Types of Techniques}
In this SMS, the techniques used in primary studies fall into two categories, i.e., Machine Learning (ML) and Natural Language Processing (NLP), according to the pilot selection process. 

\begin{figure}[h]
  \centering
  \includegraphics[width=\linewidth]{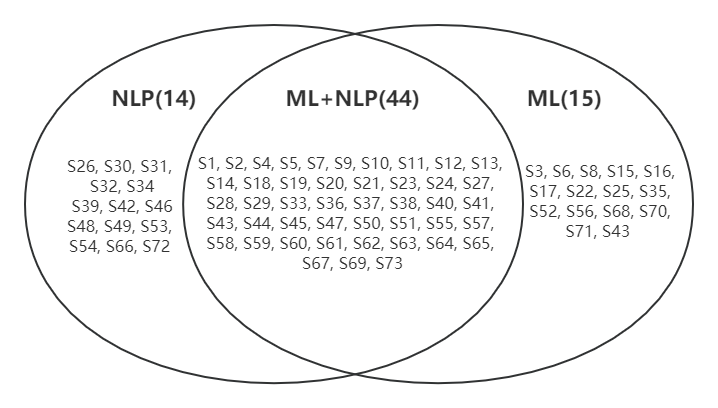}
  \caption{Venn diagram of the distribution of the selected studies over technique types.}
  \label{techtype}
\end{figure}

Figure~\ref{techtype} shows the distribution of the selected studies over technique types. It is observed that 15 studies only used ML algorithms and 14 studies only employed NLP methods. Around 59\% of the 73 selected studies used both ML and NLP for automated requirements elicitation and analysis of mobile apps.

Furthermore, we found that the 73 included studies reported 44 types of ML algorithms and 17 types of NLP techniques. Table~\ref{MLtop5} and Table~\ref{NLPtop5} show the most frequently-used (top five) ML and NLP techniques reported in the primary studies, respectively. 

\begin{table*}[h]
  \caption{The most frequently-used ML techniques (Top 5).}\label{tbthree}
  \label{MLtop5}
  \begin{tabular}{ccp{280pt}}
    \toprule
    ML Algorithm & No. of Studies & Studies\\
    \midrule
     Naive Bayes & 22  &  S3, S4, S6, S7, S8, S9, S11, S12, S18, S20, S28, S31, S22, S40, S45, S51, S55, S59, S60, S14, S15, S71 \\
     Support Vector Machine & 16  & S3, S4, S6, S7, S8, S18, S20, S28, S22, S36, S40, S45, S47, S56, S60, S71 \\
     LDA & 15 & S6, S12, S19, S20, S23, S27, S22, S40, S44, S47, S51, S56, S60, S61, S62\\
     Random Forest & 14 & S3, S7, S15, S16, S17, S24, S22, S35, S47, S50, S60, S70, S71, S73 \\
     Logistic Regression & 7 & S4, S11, S18, S20, S28, S71, S73 \\
    \bottomrule
  \end{tabular}
\end{table*}

\begin{table*}[h]
  \caption{The most frequent NLP techniques (Top 5).}\label{tbfour}
  \label{NLPtop5}
  \begin{tabular}{ccp{280pt}}
    \toprule
    NLP technique & No. of Studies & Studies\\
    \midrule
     text pre-processing techniques & 51 &  S1, S2, S4, S5, S9, S10, S11, S12, S13, S18, S19, S20, S21, S23, S24, S26, S27, S28, S30, S31, S32, S33, S34, S36, S37, S38, S39, S40, S42, S44, S45, S46, S47, S48, S49, S50, S51, S52, S53, S54, S55, S56, S58, S61, S62, S64, S65, S66, S67, S69, S72 \\
     
     TF-IDF & 14 & S2, S7, S26, S27, S31, S37, S45, S46, S54, S58, S59, S60, S65, S73 \\
   
     Bag-Of-Words & 10 & S4, S9, S11, S14, S31, S36, S48, S58, S59, S73\\
     semantic similarity algorithm  & 7  & S40, S44, S50, S51, S57, S61, S72 \\
 
     Word Vector Model (Word2vec) & 6 & S11, S37, S51, S57, S69, S72 \\
    \bottomrule
  \end{tabular}
\end{table*}

As shown in Table ~\ref{tbthree}, the most frequently used ML algorithm is Naive Bayes, which is reported in around 30\% of the 73 selected studies. Support Vector Machine, LDA (Latent Dirichlet Allocation), and Random Forest are also employed frequently in automated requirements elicitation and analysis of mobile apps, and they are reported in 16, 15, and 14 studies, respectively. Moreover, we found that many studies reported more than one ML techniques. For example, Naive Bayes, Support Vector Machine, and Random Forest were all mentioned in S3. In S36, the researchers compared the performance of four ML algorithms, i.e., J48, Naive Bayes, K-Star, and Random Forest. 

Regarding the top five NLP techniques listed in Table ~\ref{tbfour}, we observed that text pre-processing techniques are the most in-demand one, covering 51 out of the 73 selected studies. Besides, TF-IDF (Term Frequency – Inverse Document Frequency) and Bag-Of-Words also got widely used in 14 and 10 studies respectively. Similarly, multiple NLP techniques often collaborate for the RE purpose of mobile apps. For example, S2 employed both text pre-processing techniques and TF-IDF
 to mining the internal information of the review text and constructing the database for the analysis. Plus, NLP techniques often work together with ML techniques. For example, S55 used TF-IDF for text processing work and then compared the performance of four ML algorithms (i.e., Naïve Bayes, Bagging, J48, and KNN) in the automatic classification of app reviews. 

%, the researchers first processed the app dataset with a specified NLP technique called OpenNLP, and then  first processing the dataset using the NLP techniques provided by OpenNLP, the researchers directly used manual methods to further analyze the processed dataset.

%In the group with ML and NLP techniques,the comprehensive approach was used in the experiments.In [S55],  Naïve Bayes was found to be more accurate for categorizing app reviews.

%In the group with only NLP techniques,for example,in [S42],after processing the dataset using the NLP techniques provided by OpenNLP, the researchers directly used manual methods to further analyze the processed dataset.

\subsubsection{Novelty of Techniques}
To facilitate automated requirements elicitation and analysis, the researchers and/or practitioners reported two ways of employing existing automatic techniques for various RE purposes. One way is to reuse the existing techniques directly, the other one is to propose a novel method based on some existing techniques. 

Figure 6 shows the distribution of the 71 selected studies that explicitly addressed specific techniques over the novelty of techniques. We found that around 70\% of the 71 selected studies (49 studies) simply reused the existing automatic techniques, and 24 studies improved, extended, or combined the existing techniques to propose novel methods for automated requirements elicitation and analysis of mobile apps. 

\begin{figure}[h]
  \centering
  \includegraphics[width=\linewidth]{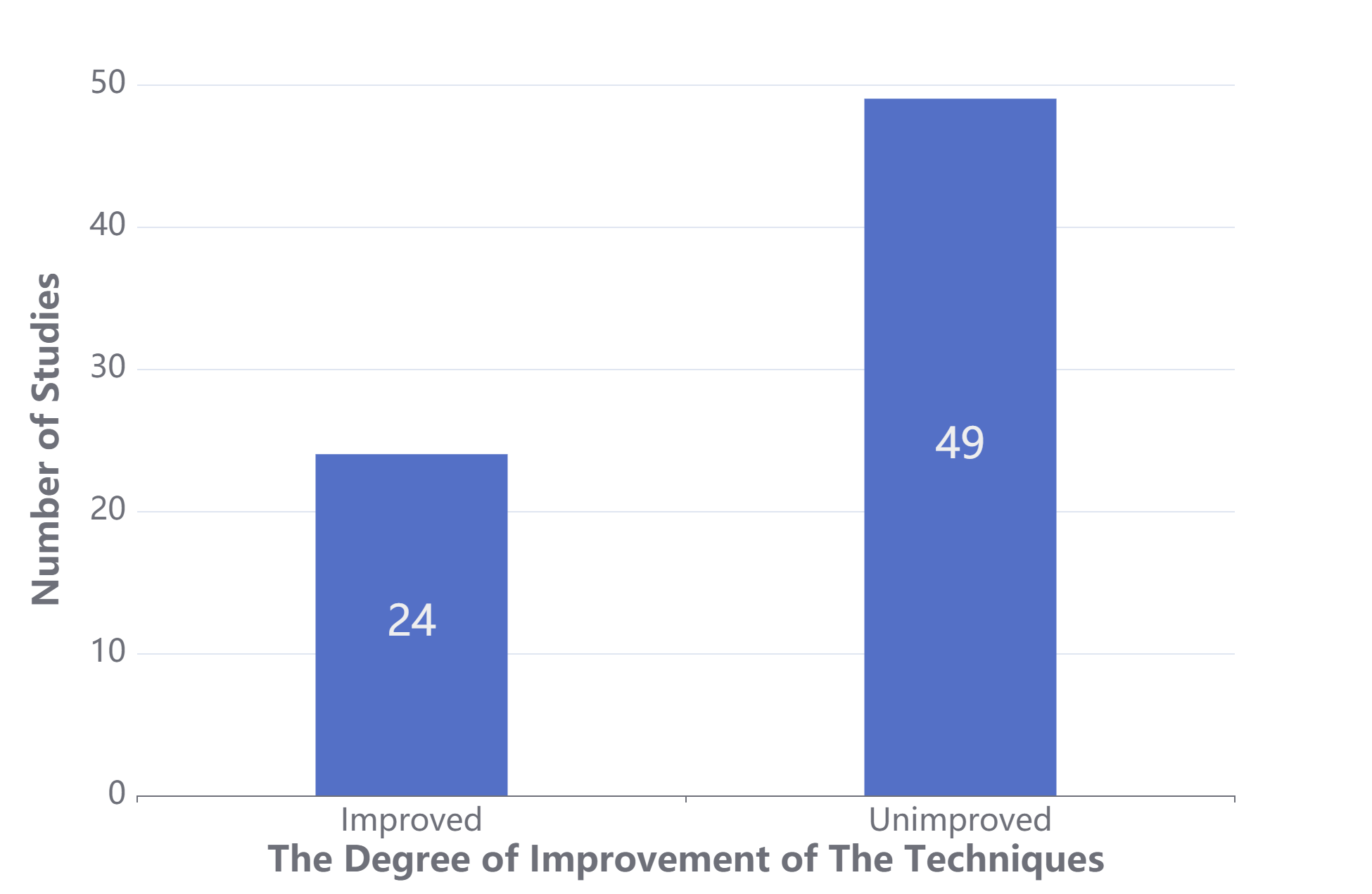}
  \caption{Distribution of the selected studies over the novelty of techniques. }
\end{figure}

% This indicates that the existing rich variety of techniques can basically meet the needs of the study.
Furthermore, the 24 studies reported two ways to improve the existing automatic techniques for the RE purpose of mobile apps. On one hand, the existing automatic techniques were extended or improved (e.g., importing new artifacts) for specified cases, which were reported in 10 studies. For example, the classical LDA modes was improved as AOLDA in S30, Entity-LDA in S38, O-LDA in S50, and APP-LDA in S64. Also, SVM was improved as hSVM in S17 to  extract detailed defect descriptions from user reviews at the sentence level. On the other hand, the improvement can be implemented by combining the different existing techniques to construct a new framework for particular RE purposes of mobile apps, and this type of improvements occurs in 14 out of 73 selected studies. In S47, for example, SAFER was proposed as a new approach consisting of a topic model and a functional aggregation recommendation algorithm. Similarly, S66 proposed a prototype system supported by MARA to extract user requirements, by combining multiple techniques.

%In the group of improvements, the improvements can be divided into two types, the first one refers to the steps of input, output or intermediate processing of existing algorithms. For LDA models,AOLDA [S30],Entity-LDA [S38],O-LDA [S50] and APP-LDA [S64],four improved LDA models were applied in the experiments. For the improvement of SVM, hSVM [S17] was obtained.

\subsection{Answer to RQ2}\label{6.3}
\begin{figure}[h]
  \centering
  \includegraphics[width=\linewidth]{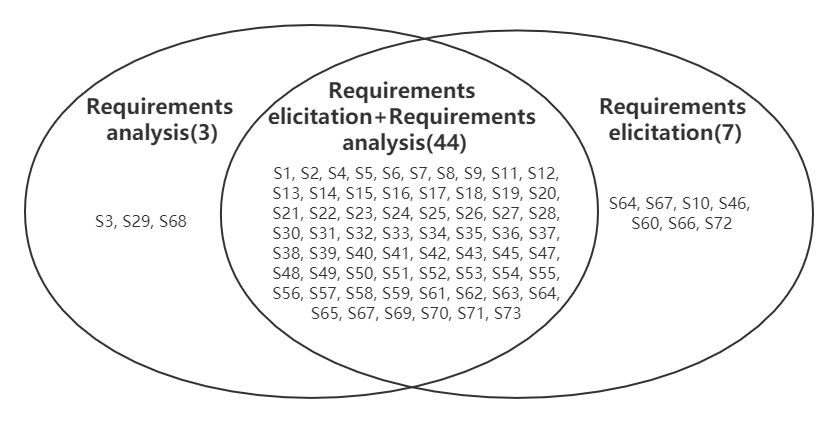}
  \caption{Distribution of the selected studies over two RE activities. }
  \label{REactivity}
\end{figure}

Figure~\ref{REactivity} shows the distribution of the selected studies over two RE activities, i.e., requirements elicitation and analysis. We found that around 86.3\% of the 73 selected studies reported both of these two RE activities. Whereas, three studies only mentioned requirements elicitation, and seven studies only aimed to requirements analysis. 

\begin{figure}[h]
  \centering
  \includegraphics[width=\linewidth]{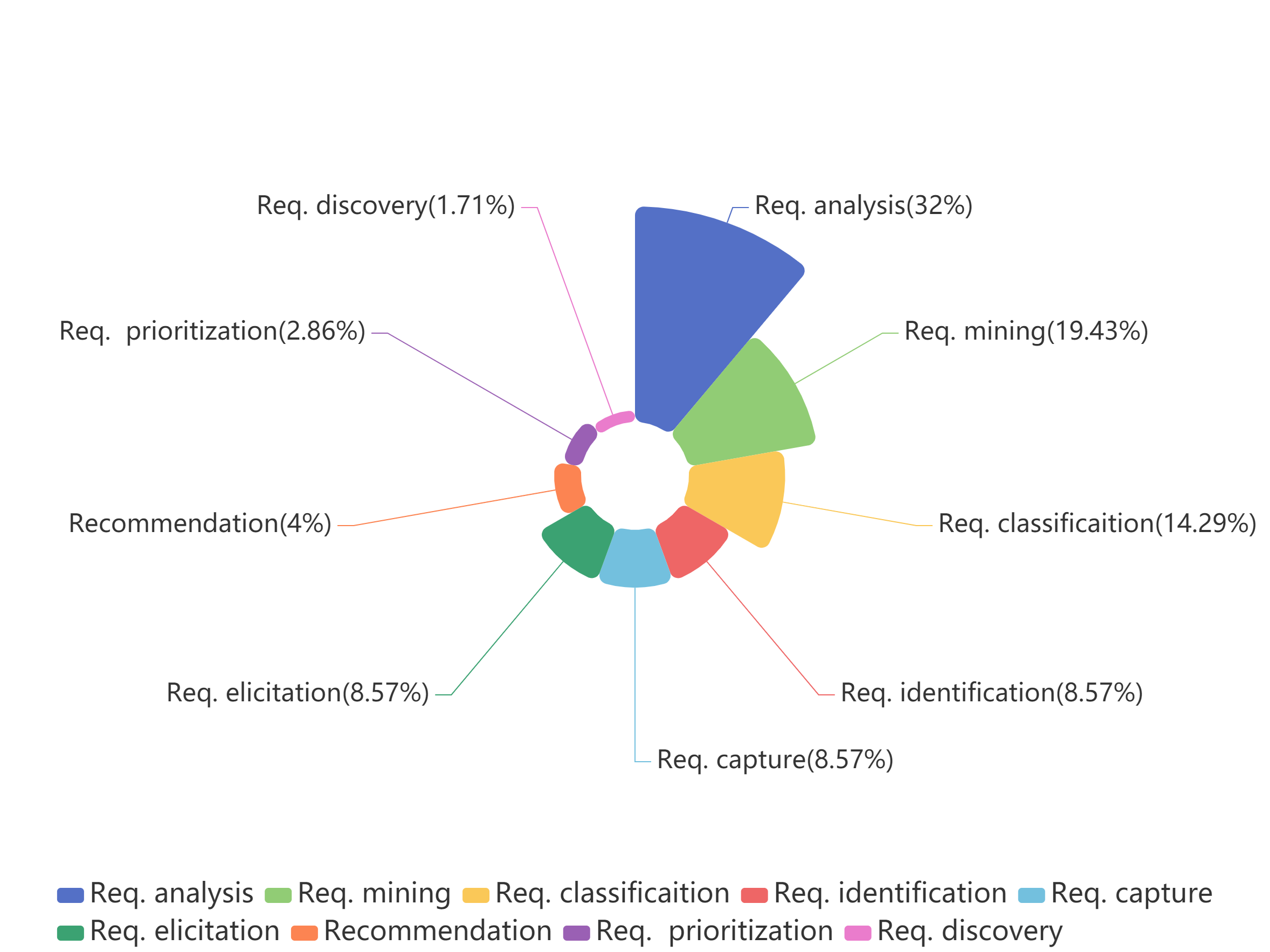}
  \caption{Distribution of the selected studies over RE tasks. }
  \label{REtasks}
\end{figure}

By zooming in on the concrete RE tasks in these two RE activities in Figure~\ref{REtasks}, the 73 selected studies reported nine RE tasks. More specifically, requirements identification (15 studies), requirements capture (15 studies), requirements mining (34 studies), requirements elicitation (15 studies), and requirements discovery (3 studies) are the five RE tasks reported to support automated requirements elicitation activity. Whereas, requirements prioritization (5 studies), requirements analysis (56 studies), requirements classification (25 studies), and requirements recommendation (7 studies) are the four RE tasks covered by the studies to support automated requirements analysis activity. In addition, we found that requirements analysis is the most identified RE task in the 73 selected studies. The following five frequently reported RE tasks are requirements mining (19.43\%), requirements classification (14.29\%), requirements identification (8.57\%), requirements capture (8.57\%), and requirements elicitation (8.57\%). Note that when identifying RE tasks from the selected studies, some studies only mentioned requirements elicitation or requirements analysis, rather than addressing concrete RE tasks. Therefore, in this subsection, requirements elicitation and requirements analysis are treated as two RE tasks.

\subsection{Answer to RQ3}\label{6.4}
Researchers use various tools to implement the techniques for data collection, text pre-processing, requirements elicitation, and requirements analysis for the RE purpose of mobile apps. In this section, we have extracted the characteristics of the automatic tools from two aspects, i.e., the types (Section ~\ref{6.4.1}) and the openness (Section ~\ref{6.4two}) of the tools.

\begin{table*}
  \caption{Distribution of the selected studies over 56 reported tools and tool types.}
  \label{56tools}
  \begin{tabular}{cccp{250pt}}
    \toprule
    Purpose  &  Total no. of studies &No. of Tools & Details\\
    \midrule
    requirements analysis & 47   &22&
 WordNet [S19, S34, S37], Weka toolkit [S7, S15, S18, S15, S24, S48, S55, S59], VADER [S45,S61], System Usability Scale [S1], SURF[S57], SentiStrength [S10, S16, S19, S20, S29, S33, S37, S38, S43], Requirements Identifier and Classifier [S65], ProSuite commercial software [S49], NLTK [S51, S52, S61, S73], KEEL toolkit [S8], DCAR [S69], CoLlaborative App Permission recommendation [S26], Clarabridge's tool suite [S25], CLAP(Crowd Listener for releAse Planning) [S15, S24, S47, S48, S68], BLIA [S57], ANGEL tool [S13], lira ticket generating tool [S11], User Request Referencer (URR) [S2], UMBC tool [S40], SAFER [S47], SentiWordNet [S17], G*Power [S14]  \\
  
     requirements elicitation & 24  &14& AR-miner [S18], SentiWordNet [S39], SAFER [S47], User review sniffing tool [S45], App Feature Extractor [S47], appFigures TOOL [S25], ARdoc tool [S22, S31, S43], LibScout [S36], MAPFEAT [S12], MARA [S32, S66], NLTK [S5, S10, S11, S12, S13, S19, S21, S33], Requirements Identifier and Classifier [S65], Standford Parse [S60], WOz [S23] \\
    
     text processing & 35  & 15 & Gensim [S11] LingPipe [S32, S47], Google Compact Language Detector [S37], NLTK [S19, S26, S27, S30, S31, S33, S34, S37, S45, S50, S51, S57, S69, S71], Weka toolkit [S8], Stanford CoreNLP [S38, S54, S64, S67], Stanford Parser [S18, S37, S60], HowNet [S10], ICTCLAS [S53], Langdetect [S61], MARA [S66], NLPIR tool [S46], TEXTBLOB [S31], table lookup [S62], Stanford POS Tagger [S39, S47]\\

     data collection & 24    & 5&Web crawler tools [S7, S16, S19, S29, S34, S35, S36, S40, S41, S47, S58, S59, S62, S64, S67, S71, S72, S64], User Reviews Extractor [S65], PlayDrone [S37], MARA [S32, S66], Backstage tool [S3], App Data Crawler [S54] \\
    \bottomrule
  \end{tabular}
\end{table*}

\subsubsection{Types of Tools}\label{6.4.1}
In this SMS, the tools used for the automated requirements elicitation and analysis of mobile apps were categorized into four types, according to their usage. They are data collection, text pre-processing, requirements elicitation, and requirements analysis. Usually, a tool could play various roles in different studies, due to different RE purposes of mobile apps. Therefore, the findings in this subsection were synthesized by investigating the roles of the tools within the scope of a certain study. 

We found that 61 out of the 73 selected studies explicitly reported tools for the automated requirements elicitation and analysis, and the remaining 12 studies did not address any tools. Figure 8 shows the distribution of these 61 studies over the four types of tools. It was observed that tools for automated requirements elicitation and analysis were covered in 24 and 47 studies respectively. 35 studies reported the use of automatic tools for data collection, such as self-developed or open-sourced Web crawlers. Tools for text processing were applied in 24 studies.

Furthermore, Table~\ref{56tools} shows the details of 56 tools reported in those 60 studies over tool types. Specifically, only five tools were reported for data collection, because many studies used `Web crawler' as a general term  referring to specified APIs, such as Google APIs in S47 and iTunes Search API in S34. Regarding text pre-processing, NLTK (Natural Language Toolkit) is the most widely used tool in Python for natural semantic processing, and it was applied in 14 out of 35 studies using tools for text pre-processing. Besides, Stanford POS Tagger was also very popular to be applied in S39 and S47. As for automated requirement elicitation, NLTK was also reported as the most popular tool in eight studies. Plus, ARDOC got  attraction from three studies, i.e., S16, S22, and S43. Finally, Weka Toolkit and CLAP were identified as the two most used tools for automated requirements analysis, account for seven and six studies respectively. 

%some tools were used in two or more studies, such as VADER([S45],[S61]) and WordNet([S19],[S34],[S37]).A total of 10 studies involved sentiment analysis, using SentiStrength and SentiWordNet tools.

\begin{figure}[t]
  \centering
  \includegraphics[width=\linewidth]{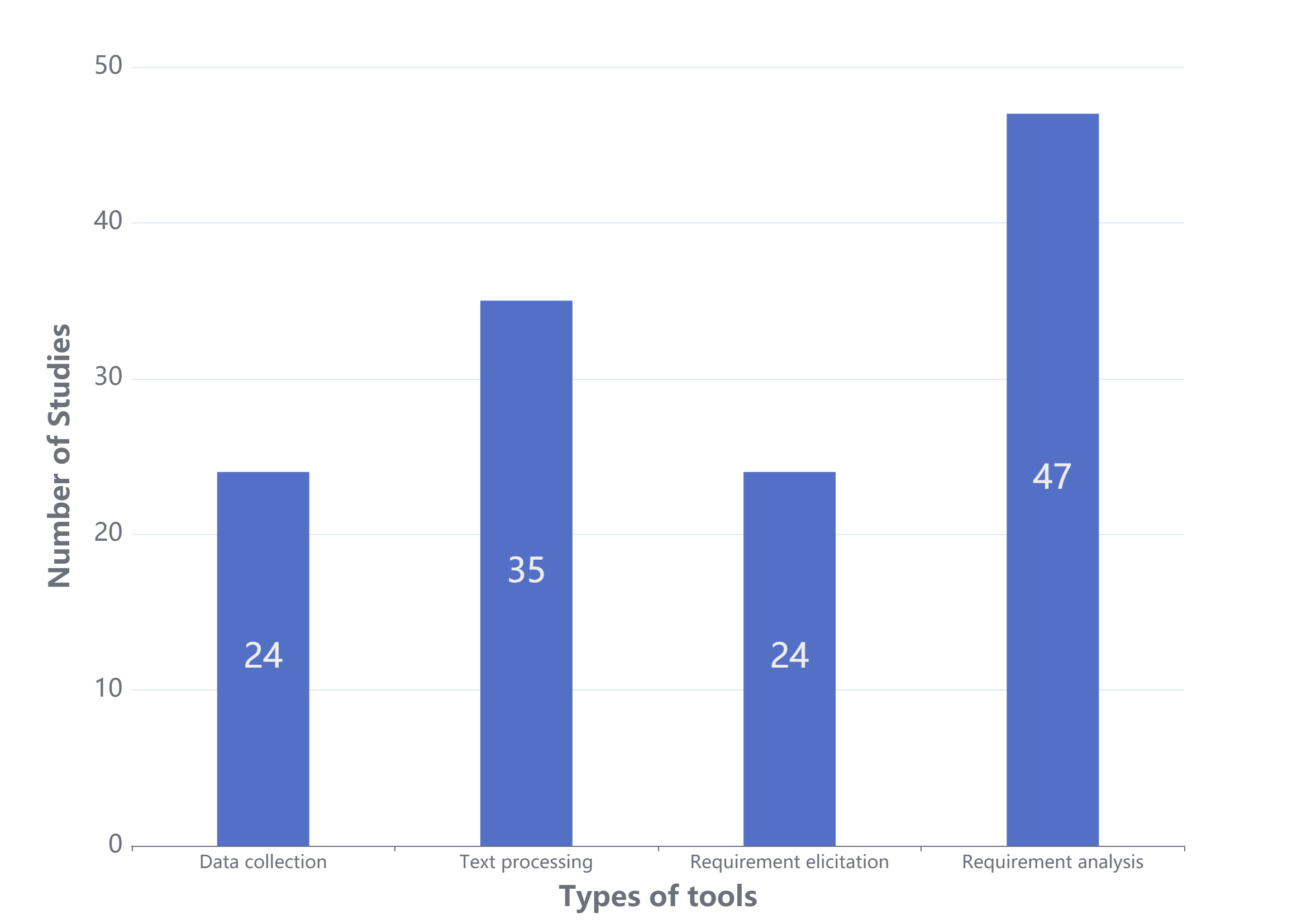}
  \caption{Distribution of 61 selected studies over tool types.}
\end{figure}

\subsubsection{The Openness of Tools}\label{6.4two}

\begin{figure}[t]
  \centering
  \includegraphics[width=\linewidth]{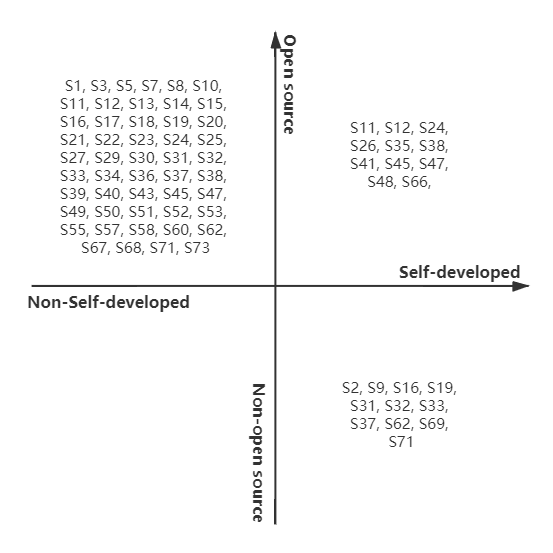}
  \caption{Distribution of the 61 studies employing tools over the two dimensions.}
  \label{twodimensions}
\end{figure}

Whether a tool is open to access and free to download is essential to evaluate the openness of the tools that implement the techniques for automated requirements elicitation and analysis of mobile apps. In this SMS, we evaluate the openness of the tools from two dimensions, i.e., open-sourced and self-developed.

We found that 12 out of the 73 selected studies did not explicitly use tools for the RE purpose of mobile employ tools. Considering the remaining 61 studies, most of them employed more than one tools for the automated requirements elicitation and analysis of mobile apps. For example, In S15, researchers use Weka Toolkit to process sentiment analysis task and use AR-miner to filter out non-informative reviews in the dataset. In S33 ,researchers use the collocation finding algorithm provided by the NLTK toolkit for extracting features from the user reviews and use the lexical sentiment analysis tool (i.e., SentiStrength) for finding users' opinions and experiences concerning features.

Figure~\ref{twodimensions} shows the distribution of the 61 studies employing tools over the aforementioned two dimensions. It was reported that 50 out of the 73 included studies use open-sourced tools provided by other organizations or individuals. 14 studies employed the open-sourced tools developed by the authors of these studies. Whereas, eight studies developed tools for the RE purpose of mobile apps, but these tools are not free-available.

%The open source nature of the tools used to conduct the study is also an aspect of our research, and the open source nature of the tools can be divided into open source and non-open source, self-developed and non-self-developed. There should be a combination of four cases, because non-self-developed and non-open source are not possible cases. So there are three actual, and we will count all the tools involved in the study. Figure 10 shows the results of the openness of the tools.

% The total number of open source and non-self-developed tools is 50, accounting for 69.4$\%$. The important components of this part of the tool are the results of previous researchers, open source software and executable code.non-open source and self-developed has only 8 items, accounting for 11.1$\%$. In this research area, most of the tools are highly shared.

\section{Discussion and Implications}
This section presents our reflection on the demographics of the 73 selected studies as well as the answers to the three RQs.

\subsection{Overview of Primary Studies}
This SMS clearly indicates that the automated requirements elicitation and analysis of mobile apps is a rising research area for researchers and practitioners. We concluded this from the growth of the published literature (Figure ~\ref{year}) in the past five years. Especially, the total number of primary studies published between 2017 and 2019 account for 64.4\% of the 73 selected studies. Moreover, 85.0\% of the 73 selected studies are publish in international conferences or workshops. The reason could be that with the growing numbers of mobile apps and their app data, automated techniques, methods, and tools are becoming popular techniques for processing and analyzing the massive app data. Surprisingly, we noticed that the number of primary studies rapidly decreases in 2020. One possible reason could be that the world-wide explosion of COVID-19 did greatly affect the academia.

%aims to investigate the state-of-the-art of the techniques and tools for automated requirements elicitation and analysis of mobile apps. selected a total of 73 relevant papers from 2012 to 2020 for the literature survey, covering This is an important study in this area. As the number of years increases, the number of studies in this area also shows an increasing trend. There is a strong upward trend in 2016-2017, peaking in 2019, with the three years 2017-2019 accounting for 64.4$\%$ of all studies. Therefore, we can conclude that research in this field is increasing and showing a hot trend. However, at the same time, we note that in 2020, the number of studies suddenly decreases. One possible explanation is related to the COVID-19 epidemic which is not conducive to the conduct of academic research.

\subsection{Characteristics of Techniques (RQ1)}

First, we concluded that it is still necessary to involve human beings in the automated requirements elicitation and analysis of mobile apps, because more than half of the 71 studies reporting automatic techniques were conducted with the supplement of manual tasks, including polling, surveys, manual labelling, evaluation, etc. One reason could be that supervised ML methods are very popular (as shown in Table ~\ref{tbthree}), and a necessary truth set should be constructed by manual labelling. Another reason could be that the accuracy of employing pure automatic techniques is often much lower than that by manual elicitation and analysis of requirements. Although the researchers continue to improve the accuracy, it is quite difficult, or even impossible, to get the perfect 100\%. Therefore, it implies that for the automated requirements elicitation and analysis of mobile apps, semi-automated techniques are more suitable than pure automatic techniques. Also, due to time and human cost of manual tasks, it is expecting to explore more effective techniques with less involvement of human being for automated RE activities of mobile apps.

%With regard to techniques, we found that 71 out of 73 articles used techniques. 13 more semi-automatic techniques than fully automatic ones. Most of the semi-automated techniques increased compared to fully automated ones in terms of human evaluation of study results or human participation in the study as a source of data sets (e.g., polling, surveys, manual annotation). 

% Most of the automated techniques are single purpose or a combination of multiple fixed purposes, and the programmed techniques are very precise. However, purely automatic methods are of limited use in the face of unquantifiable situations. Therefore, in the case of comprehensive research analysis, researchers often use semi-automatic methods to complement the shortcomings of purely automatic methods with reliable manual analysis. This is the reason why researchers prefer manual methods as a complement to automatic methods.

Regarding the types of techniques, it was concluded that the combination of NLP and ML techniques is the most popular way, covering 59.0\% of primary studies, to support automated requirements elicitation and analysis of mobile apps. This reveals that first, these two RE activities of mobile apps are facing new challenges so that individual techniques are insufficient to provide decent solutions. Second, the combination of multiple efficient methods or algorithms is becoming more and more popular, and got accepted or has been adopted by some researchers and practitioners. The finding in Section ~\ref{6.2} also signals a question that why NLP and its collaboration with ML were the most reported techniques in our SMS, accounting for around 80.0\% of the primary studies. The reason could be that current app datasets mainly consist of textual data, and this meets the usage of either NLP or ML. Therefore, we argue that more combination of separate techniques can be proposed to process new types of app data and to improve the automation degree of those techniques. 

%has been accepted and adopted by many   accounts for 60.6$\%$ of all techniques. This is because when processing native datasets, the complete process often requires NLP techniques to process the dataset first, and only after obtaining the processed dataset can ML be used for classification or analysis. Most of the studies that involve only a single domain of techniques only cover one aspect of the complete process. For example,in [S52],the dataset used is a processed set of mobile application reviews obtained from an open source platform

Considering the novelty of techniques, around 70\% of the selected studies preferred to reuse existing techniques for automated requirements elicitation and analysis of mobile apps, rather than proposing a new one for the specific case. The reason could be that the direct reuse of automatic techniques may provide good enough solutions to simple cases. However, it is exciting that some researchers or practitioners, especially the authors of those 24 studies, have explored challenging cases on this research topic with newly proposed methods. This indicates that our research area may be more active, due to the success of these 24 studies with improved techniques. In addition, it is necessary to encourage more exploration on the improvement, extension, and combination of existing techniques, in order to support automated requirements elicitation and analysis of mobile apps.

Actually, the weakness of using automatic techniques always exist. For example, in most cases, the automatic techniques cannot be directly reused for specified RE tasks or in specified research dataset. Plus, it is inevitable that the accuracy of pure automatic techniques is relatively lower than that of manual efforts. Nevertheless, this SMS reported that automatic techniques have caught the eyes of researchers and/or practitioners. The reason is that for the research on automated requirements elicitation ans analysis, the size of research dataset keeps growing to cover millions of app data. Since pure manual efforts are impossible, automatic techniques become a `Not perfect but good enough' solution for our research topic. 
Then, manual efforts, such as questionnaires and manual labeling, are encouraged as supplement to pure automatic techniques. This is also the reason why the proportion of semi-automatic techniques is greater than that of pure automatic techniques.
%Reviewing Figure 8, we can see that the improved techniques are relatively few and are mainly ML-oriented techniques. This is probably because the ML-techniques are used in a wider range and are more focused on the clustering and classification of the processed dataset, compared to the NLP-techniques, which are focused on the pre-processing of the dataset and have a greater direct impact on the research results, and contain more parameters, so it is easier to optimize and improve the effects of the techniques from multiple perspectives. It is easier to optimize and improve the effect of technique from multiple perspectives.

\subsection{RE Tasks (RQ2)}

The findings in Section ~\ref{6.3} indicate that most primary studies (44 out of 73 studies 60.2\%) report both requirements elicitation ans analysis activities. The reason could be that these two RE activities are logically related, and requirements elicitation always generate inputs to requirements analysis.

Moreover, five RE tasks were identified for requirements elicitation, and four RE tasks were reported to support requirements analysis. Specifically, requirements analysis, requirements mining, and requirements classification are the top 3 reported RE tasks. The reason why requirements mining is the second most important RE tasks could be that it is the basis of further study on requirements. The rank of requirements classification indicates that this RE task help filter specific requirements from the requirements pool, according to certain requirements types.

%Reviewing Figure 9, we can conclude that the two requirements activities, requirements analysis and requirements acquisition, are basically performed simultaneously. This is because after acquiring the native dataset, the requirements need to be extracted before the requirements can be analyzed, a process that is consistent with the general logic of our problem analysis. Requirements analysis is a collection of many requirements activities, including requirements classification, requirements prioritization, requirements negotiation, and many other activities. This is consistent with the conclusion that a large number of ML techniques are used for classification, as we found in Figure 6 that ML techniques occupy the most important part.

\subsection{Characteristics of Tools (RQ3)}

The findings in Section ~\ref{6.4} report four types of tools used for automated requirements elicitation and analysis of mobile apps. We conclude that the largest group of tools aim to support requirements analysis. This indicates that the importance of the requirements analysis activity, which is consistent with the findings in Section ~\ref{6.3}. Text processing is the tool type with the second largest number of primary studies. This indicates the domination of textual information in app datasets. 

Furthermore, our SMS reports 56 tools used in the 73 selected studies. We conclude that a certain tool can be used not only in different studies but for different purposes, such as NLTK. This indicates that unlike techniques, powerful tools can integrate several functions, and then work in different stages of the solutions. In addition, we found that the number of tools used in requirements elicitation is around half of that in requirements analysis. The reason could be that more manual efforts are needed to capture requirements, due to the common understanding on the nature of requirements.

In addition, our SMS reviewed the reported 56 tools from two perspectives, i.e., open-sourced and self-developed. 62 out of 73 selected studies employed open-sourced tools. The tools used in the 51 out of these 62 studies are developed and maintained by other organizations or companies, rather than the authors of the 51 studies. This indicates that open-accessed and free-to-download tools are always welcome, especially in automated requirements elicitation and analysis of mobile apps. One reason could be that the researchers and practitioners mainly concentrate on the novelty of research methods, rather than tool development. Another reason could be that some tools have been developed for common purposes, e.g., data collection and text processing, and reuse is the easiest and fastest way to support automatic requirements elicitation and analysis of app datasets. 

Finally, we notice that integrated frameworks and tools have been used more and more frequently in recent years. Taking the CLAP framework as an example, it can be applied in the whole process starting from text pre-processing to requirements elicitation, and then to requirements analysis. Actually, most of these integrated frameworks and tools are combinations of existing separate tools, rather than novel tools. This also implies that the separate tools components have been well developed to effectively support individual RE activities. We therefore think that from the RE activities, more investigations on how to integrate existing tools as well as improve or extend the popular tools for different RE tasks or purposes. Whereas, from the perspective of tools, it is necessary to develop more tools to meet the context of other RE activities and/or data from other sources.

%focus should be on developing tools for new requirements activities or new types of datasets, rather than duplicating existing requirements activities tools.

%In general, as the field becomes more and more popular, related techniques and tools are gradually improved, and the existing techniques and tools can meet the needs of research in general, but for the problems of customization and innovation, The conventional techniques and tools are not effective and require innovative development and improvement. The problem of homogenization of research, i.e., researchers focus on a few fixed types of techniques and tools to deal with different mobile application requirements activity tasks. This needs our attention.

\subsection{Implications}

This SMS has some implications for RE researchers and practitioners. 

First, our results provide advice on techniques and tools for the researchers and practitioners. If they want to start or extend their work on automated requirements elicitation and analysis of mobile apps, we suggest them to try these techniques or tools in pilot study. Plus, RE researchers should notice that although automatic techniques and tools are being widely accepted and used in this research topic, manual efforts on requirements elicitation and analysis cannot be overlooked. More exploration on how to reduce human cost with automatic techniques and tools should be conducted by both researchers and practitioners.   

Second, openness is the main factor to determine whether a certain tool can be adopted in the automated requirements elicitation and analysis of mobile apps. This is consistent with user's demands in tools for the RE purpose of other types of software. For either RE researchers or practitioners, they are willing to share self-developed tools to public, in order to popularize their achievements. We therefore think that this sharing should be encouraged to call for more and more techniques and tools for the automated requirements elicitation and analysis of mobile apps. However, who is responsible for the development of common tools and how to promote the reuse of development tools in research and practice? We encourage practitioners to contribute their experience to making techniques and tools serve more RE researchers and other practitioners. 

Third, mixed techniques and integrated tools caught more eyes from RE researchers and practitioners. Therefore, we encourage researchers and practitioners to conduct more explorations on the employment or combination of techniques in other application domains. This might be a good idea to have more useful techniques and tools for the automated requirements elicitation and analysis of mobile apps.

\section{Limitations}
This SMS has some limitations. 
First, the primary studies were selected from the automatic search in two digital libraries, i.e. IEEE Xplore and Scopus. Due to limited time spent on data extraction, we skipped snow balling and additional automatic search on other portals. Would the results differ if we explore more digital databases and conduct snow balling? There is a good chance that our findings would be similar, if other digital libraries, such as Web of Science, ACM, or Wiley, are supplemented in the near future. 

Second, both the selection and data extraction of primary studies depended on our RE knowledge. To avoid the possible bias and make the survey reproducible by other researchers or practitioners, we developed a study protocol to define our search strategy and specify the study selection procedure with IC and EC. Especially, to evaluate and improve the protocol, we conducted a pilot selection, and the first two authors were involved in the three round selection to assure a common understanding on study selection. Furthermore, the first two authors performed the data extraction on the result of pilot selection and generated a template to describe the data derived from primary studies and support our qualitative synthesis.  

%Second, throughout the process of conducting paper screening, special care should be taken not to change the overall results when adjusting the research questions and key searches. The process of determining inclusion-exclusion criteria is an intensive and iterative one. After the first round of loosely setting the criteria, the obtained papers were further screened and we found that although the number of papers passing the criteria was excessive, a significant proportion of papers actually did not meet the requirements. Therefore, a number of additional iterations were performed to obtain the inclusion-exclusion criteria that better meet our requirements. It would have been better to adopt more stringent criteria at the beginning to facilitate the experiment.

Third, in practice, it is difficult to specify the types of either techniques or tools. Many tools, such as NLTK and Weka, are integrated with several techniques, so that they can work across different research for different purposes. In this paper, the tools types were specified according our experience and knowledge. Therefore, it is needed to evaluate the suggested categories of tools to support our qualitative synthesis on automatic tools.  

\section{Conclusions and Future Work}

On the basis of 73 selection publication, this SMS provided an overview on the types of techniques and tools used for the automated requirements elicitation ans analysis. Our research revealed the follows. First, more and more research employed automatic techniques and tools for requirements elicitation and analysis of mobile apps. Second, semi-automatic techniques, especially the combination of ML and NLP techniques, are the most used techniques in this research topic. Specifically, Naive Bayes and text pre-processing techniques are the most frequently used ML and NLP techniques respectively. Third, most of the techniques and tools support both requirements elicitation and analysis. Requirements analysis, mining and classification are the top three RE tasks with the support of automatic techniques and tools. Finally, the most popular tools are open-sourced and non-self-developed, and they are mainly used in requirements analysis and text processing. 

%in the area of techniques and tools for automated requirements elicitation and analysis of mobile apps.We selected a total of 73 existing studies on this topic, choosing three questions about techniques and tools for automated requirements, and analyzing each study in practice. Our research shows that the use of technologies and tools by researchers and practitioners has been on the rise in recent years, and we provide experimentally supported answers to questions about the open source nature and types of technologies and tools. We also note that the use of open-source tools obtained from others' research or the Internet can meet most of the research goals, but for seminal problems, such tools are not sufficiently applicable, while self-developed tools are more effective.

The next steps of our SMS include: (1) to add publications on this research topic from other digital libraries (e.g. ACM and Web of Science); (2) to further explore the characteristics of techniques and tools used in requirements elicitation and analysis respectively.

\bibliographystyle{ACM-Reference-Format}
\bibliography{sample-base}

%%[S1] 

\balance
\end{document}